\documentclass[12pt,preprint]{aastex}

\slugcomment{To appear in the Astrophysical Journal}

\lefthead{Skinner et al.}
\righthead{LkCa 15}

\begin{document}

\title{X-ray Irradiation of the LkCa 15 Protoplanetary Disk}

\author{Stephen L. Skinner\footnote{CASA, Univ. of Colorado,
Boulder, CO, USA 80309-0389; stephen.skinner@colorado.edu} and
Manuel  G\"{u}del\footnote{Dept. of Astronomy, Univ. of Vienna, 
T\"{u}rkenschanzstr. 17,  A-1180 Vienna, Austria; manuel.guedel@univie.ac.at}}

%
\newcommand{\ltsimeq}{\raisebox{-0.6ex}{$\,\stackrel{\raisebox{-.2ex}%
{$\textstyle<$}}{\sim}\,$}}
%
\newcommand{\gtsimeq}{\raisebox{-0.6ex}{$\,\stackrel{\raisebox{-.2ex}%
{$\textstyle>$}}{\sim}\,$}}

\begin{abstract}
LkCa 15 in the Taurus star-forming region has recently gained attention as the  
first accreting T Tauri star likely to host a young protoplanet. High spatial 
resolution infrared observations have detected the suspected protoplanet within a dust-depleted 
inner gap of the  LkCa 15 transition disk at a  distance of $\sim$15 AU from the star. 
If this object's status as a protoplanet is confirmed, LkCa 15 will serve as a
unique  laboratory for constraining physical conditions within a planet-forming disk.
Previous models of the LkCa 15 disk have accounted for disk heating by the
stellar photosphere but have ignored the potential importance of X-ray ionization 
and heating. We report here  the detection of LkCa 15 as a bright 
X-ray source with {\em Chandra}. The X-ray emission is characterized by a 
cool heavily-absorbed plasma component at kT$_{cool}$ $\approx$ 0.3 keV and a 
harder component at kT$_{hot}$ $\approx$ 5 keV. 
We use the observed X-ray  properties to provide initial estimates of the 
X-ray ionization and  heating rates within the tenuous inner disk. These
estimates and the observed X-ray properties of LkCa 15 can be
used as a starting point for developing more realistic disk models of 
this benchmark system.
\end{abstract}





\section{Introduction}
In a recent groundbreaking discovery, Kraus \& Ireland (2012, hereafter KI12)
have reported the direct detection of what is likely the first
exoplanet orbiting a young T Tauri star (TTS). Infrared images obtained using a 
novel masked aperture interferometry technique revealed the suspected 
protoplanet  known as LkCa 15b orbiting the solar-mass TTS LkCa 15 
in Taurus. The protoplanet  is offset $\approx$80 mas from the 
star and the  separation accounting for projection effects is 
estimated to be 15.9 $\pm$ 2.1 AU (KI12). The protoplanet is located within a
dust-depleted  cavity in the LkCa 15 inner  disk. The estimated
mass of the protoplanet is M$_{p}$ $\sim$ 6 M$_{Jup.}$ at an assumed age
of $\sim$1 Myr. The blue color and rather high luminosity
L$_{p}$ $\sim$ 10$^{-3}$ L$_{\odot}$ of LkCa 15b suggest that it
may still be accreting. The large radius of the  cavity
in which it is located (R$_{cav}$ $\approx$ 50 AU) is a subtle hint
that other as yet undetected protoplanets may  be present
(Andrews et al. 20011a,b; hereafter A11a,b). If the status of LkCa 15b as
a protoplanet is confirmed by further observations, this young
star$+$exoplanet system will provide much-needed observational 
constraints on the physical conditions in disks that are actively
forming planets and will stringently test planet formation models.

LkCa 15 has been extensively studied at optical, IR, and (sub)millimeter
wavelengths. Its properties are summarized in Table 1.
Interferometer mm/sub-mm dust continuum observations (A11b;
Isella et al. 2012) have shown that dust is heavily depleted
inside a radius of $\sim$50 AU, but a small amount of residual
dust may remain (A11a,b). CO observations reveal that gas
is still present in the dust-depleted cavity (Pi\'{e}tu et al. 2007).
Such a  dust-depleted inner disk with the absence of a
near-IR excess is frequently referred to as a  {\em transition disk},   
but different definitions of this term have emerged in
the literature as reviewed by Williams \& Cieza (2011).

Several possible mechanisms for clearing the large cavity were examined
by A11a, who  concluded that dynamical interactions with  a low-mass companion -
either a brown dwarf or giant planet on a long-period orbit -
are the most likely explanation. However, questions remain as to whether
such a large gap could have been cleared by a {\em single} planet,
suggesting (indirectly) that a planetary {\em system} may be present.
But other explanations for the large disk cavity have been proposed including
radial variations in grain size and dust opacity (Isella et al. 2012). 

Previous models of the LkCa 15  disk have considered 
photospheric heating of the disk by the central star, but have
neglected X-ray heating. For example, the recent study of 
disks based on high resolution Submillimeter Array (SMA) observations by
A11a quoted a 3$\sigma$  upper limit on the X-ray luminosity of LkCa 15
of log L$_{x}$ $<$ 29.6 ergs s$^{-1}$ based on its non-detection
in the {\em ROSAT} All-Sky Survey (Neuh\"{a}user et al. 1995).
We report here the first detection of LkCa 15 as a bright X-ray
source with an intrinsic (unabsorbed) X-ray luminosity  of at
least log L$_{x}$ $\approx$ 30.4 ergs s$^{-1}$, significantly higher
than the  {\em ROSAT} upper limit cited above. Our primary objective here is to
elucidate the X-ray properties of LkCa 15 so that X-ray effects
can be accounted for in more refined disk models. 
We  provide initial estimates of X-ray ionization and
heating rates based on the observed X-ray properties and  
previously published disk models. 

We emphasize that strong X-ray (and EUV) emission from the central 
star has important consequences for
disk evolution, disk chemistry, accretion, mass-loss, and planet formation. X-rays 
ionize and heat  material  in the disk (especially in the  outer layers),
the wind, and in the  outer atmospheres of any protoplanets 
in the inner disk region near the star. Increased ionization of
disk gas by X-rays strengthens the coupling of the disk to the
stellar magnetic field and thereby influences angular momentum
transport (Balbus \& Hawley 1991). X-rays destroy dust  near the star and 
X-ray induced photoevaporation increases mass-loss  and  can lead to 
formation of inner disk holes that affect the accretion rate 
(Owen et al. 2010).

\section{Chandra Observation}

The {\em Chandra} observation (ObsId 10999) was carried out on
27 December 2009 from 09:40 - 12:54 TT with an exposure live time
of 9813 s.  Exposures were obtained using the ACIS-S (Advanced CCD 
Imaging Spectrometer) array in VFAINT  timed-event mode 
with  3.2 s frame times. LkCa 15  was placed at the nominal
aimpoint on the ACIS-S3 CCD (Table 2).  For an on-axis point source,
the ACIS-S 70\%  encircled energy radius at 2 keV  is
R$_{70}$ $\approx$ 1.$''$17 and the 90\% encircled energy  
radius  is R$_{90}$ $\approx$ 1.$''$96. 
Further information on {\em Chandra} and its instrumentation can 
be found in the {\em Chandra} Proposer's 
Observatory Guide (POG)\footnote {See http://asc.harvard.edu/proposer/POG}.

The pipeline-processed data  files provided by the {\em Chandra} X-ray
Center (CXC) were  analyzed using standard science
threads with CIAO version 4.4\footnote{Further information on 
{\em Chandra} Interactive
Analysis of Observations (CIAO) software can be found at
http://asc.harvard.edu/ciao.}.
The CIAO processing  used recent calibration
data from CALDB version 4.4.10.
Source events, spectra, and light curves were extracted from a circular region of
radius 1$''$.5 (3 ACIS pixels) centered on the X-ray peak.
Background was extracted from a source-free  annulus centered on LkCa 15
with inner and outer radii of 5$''$ and 15$''$. Background is negligible,
amounting to less than 1 count (0.2 - 8 keV) within the $r$ = 1$''$.5 extraction
circle during the 9.8 ks exposure.
CIAO {\em specextract} was used to extract   
spectra along with source-specific
response matrix files (RMFs) and auxiliary response files (ARFs).
Spectral fitting, timing analysis, and image analysis were undertaken with the HEASOFT 
{\em Xanadu}\footnote{http://heasarc.gsfc.nasa.gov/docs/xanadu/xanadu.html.}
software package including XSPEC vers. 12.7.1. XRONOS vers. 5.2.1, and XIMAGE vers. 4.4.
Additional tests  for source variabilility
were carried out on  energy-filtered source event lists 
using $\chi^2$ statistics and the Bayesian-method CIAO tool {\em glvary} 
(Gregory \& Loredo 1992, 1996).

\section{Results}

Table 2 summarizes the basic X-ray properties of LkCa 15.
The measured X-ray centroid of the source is offset by only
0$''$.08 from the {\em HST} GSC v2.3.2 position of LkCa 15 (Fig. 1).
This small offset is well within {\em Chandra}'s ACIS-S absolute
astrometric accuracy of $\approx$0.$''$42 (90\% confidence) 
\footnote{http://cxc.harvard.edu/cal/ASPECT/celmon/ }. There are
no other sources in the {\em HST} GSC or 2MASS catalogs within 10$''$ of
LkCa 15, giving high confidence that LkCa 15 is the X-ray source.
There is no evidence to date that LkCa 15 is a binary star (Nguyen et al. 2012).
The {\em Chandra} ACIS-S PSF core has FWHM $\approx$ 0.$''$74 at 1.5 keV
so the  angular resolution is not sufficient to determine if any of
the detected emission originates in the protoplanet itself.
The CIAO tool $srcextent$ gives an observed source size of
0.$''$46 [0.$''$43 - 0.$''$49, 90\% confidence], so there is no 
indication that the source is extended. We assume below that all
of the X-ray emission is stellar.

The CIAO tool $glvary$ gives a probability of constant count rate
P(const) = 0.91 using source events in the 0.2 - 8 keV range.
Chi-squared analysis  of binned light curves constructed from events in
the 0.2 - 8 keV range gives P(const) = 0.68 (200 s bins) and
P(const) = 0.54 (400 s bins). Thus, we find no compelling evidence
for variability in LkCa 15 during the short 9.8 ks ($\approx$2.7 hr) exposure, 
but further time monitoring would be useful to determine if large X-ray
flares occur, as is common amongst TTS. Such flares, if present, 
usually harden the X-ray spectrum and  bombard the inner disk with
high particle fluxes.

Figure 2 shows the ACIS-S CCD spectrum of LkCa 15. Low-energy absorption
is clearly seen at energies E $<$ 0.6 keV. Most of the detected emission
lies in the energy range 0.5 - 2 keV but harder emission is present.
A total of 72 out of 592 source events have energies E $>$ 2 keV.
Spectral fits with an absorbed thermal plasma model require
at least two temperature (2T) components. Table 3 compares the fit 
results obtained with  three different 2T models. The model fits
were obtained using the Astrophysical Plasma Emission Code $apec$ 
and its variable-abundance version $vapec$ as implemented in 
XSPEC (Smith et al. 2001). Model A is a
solar abundance  2T $apec$ optically thin plasma model. Model B
is similar except that the global metallicity $Z$ was allowed to vary
and it converged to a best-fit value Z = 0.46 Z$_{\odot}$, but
the value is not tightly-constrained.
Model C is a 2T $vapec$ model for which the 
abundances of individual elements were held fixed at values typical of
TTS in Taurus (G\"{u}del et al. 2007; Scelsi et al. 2007). 

All three models require a cool plasma component at kT$_{1}$ $\approx$ 0.2 - 0.3 keV
and a  hotter component at kT$_{2}$ $\approx$ 4 - 5  keV, with most of
the volume  emission measure as gauged by the XSPEC $norm$ parameter
residing   in the cool component. The hot
component is required to reproduce the hard emission detected
above 2 keV. The upper 90\% confidence bound on  kT$_{2}$
is not tightly constrained due to the paucity of counts
above 2 keV.

All three models in Table 3 are statistically acceptable and there is 
very little difference in their goodness-of-fit as determined by $\chi^2$ statistics. 
However, on physical grounds model C is clearly the most realistic.
It gives a best-fit absorption column density
N$_{\rm H}$ =  3.7 [2.4 - 5.1; 90\% confidence] $\times$10$^{21}$ cm$^{-2}$.
Using the N$_{\rm H}$ to A$_{\rm V}$ conversion of
Gorenstein (1975), the above N$_{\rm H}$ equates to
A$_{\rm V}$ = 1.7 [1.1 - 2.3] mag and the conversion of
Vuong et al. (2003) gives A$_{\rm V}$ = 2.3 [1.5 - 3.2] mag.
These A$_{\rm V}$ values are consistent with the range
A$_{\rm V}$ = 1.3 - 1.7 mag determined from other studies
(e.g. Espaillat et al. 2010). In contrast, the N$_{\rm H}$
values determined from models A and B are twice as large
and imply  A$_{\rm V}$ values that are much higher
than anticipated. In addition, the
ratio   log L$_{x}$/L$_{bol}$ = $-$3.18 from model C is
similar to other TTS in Taurus (Telleschi et al. 2007),
but the ratios determined from models A and B are
unusually high. We thus adopt model C as the reference
model used in the ionization and heating rate calculations
below.

\vspace*{0.5in}

\section{Discussion}

\subsection{X-ray Heating and Ionization}

X-ray photons are able to penetrate  the surface layers of  protoplanetary
disks, resulting in increased ionization and heating. In dense environments,
hard X-rays are particularly important because they suffer less absorption
and reach deeper layers in the disk. But, as shown below, the effects of 
softer X-rays cannot be ignored in low-density environments such as the
inner region of the LkCa 15 transition disk. X-ray heating is produced   
by fast electrons  ejected by atoms which become ionized after
absorbing X-ray photons. These fast electrons collide with
and heat other atoms and electrons in the disk. Additional X-ray heating 
occurs as a result of other processes as discussed by 
Glassgold et al. (2012; hereafter G12).

We provide below initial estimates of the X-ray ionization and heating 
rates in the inner disk (r $\leq$ 15 AU) of LkCa 15. We use cylindrical coordinates
(r,z) where r is the distance from the star in the disk midplane and z 
is the height above the  midplane. The disk is assumed to be azimuthally 
symmetric.  Our analysis  follows that of Glassgold et al. (1997a,
hereafter G97a; see also  Glassgold et al. 1997b), 
Igea and Glassgold (1999, hereafter IG99), Shang et al. (2002; hereafter S02), 
and Glassgold et al. (2004). We ignore scattering effects, which have
been discussed by IG99. We make use of properties of the LkCa 15
disk, especially the dust-depleted cavity  inside 
$\sim$50 AU, derived by A11a and Isella et al. (2012)
using data from (sub)millimeter interferometers.
We assume that the disk is composed of solar-abundance material,
but settling of heavier elements toward the disk midplane
could result in heavy-element depletion in the outer layers (G97a).

For a thermal X-ray spectrum with characteristic plasma temperature
T$_{\rm x}$, the ionization rate at a distance $r$ 
is (eq. [3.9] of S02):

\begin{equation}
\zeta \approx \zeta_{\rm x} \left[{ \frac{r}{R_{\rm x}}} \right]^{-2} \left[{ \frac{kT_{x}}{\epsilon_{ion}}} \right] I_{p}(\tau_{\rm x}, \xi_{0})~~~{\rm s}^{-1}
\end{equation}
where $R_{\rm x}$ is the distance of the X-ray source above (or below) the center of the disk,
$\epsilon_{ion}$ is the energy to create an ion pair ($\epsilon_{ion}$ $\approx$ 37 eV for a solar abundance
H$+$He plasma), and the function $I_{p}(\tau_{\rm x}, \xi_{0})$ is an asymptotic approximation of  
the X-ray attenuation at optical depth $\tau_{\rm x}$ 
for a specified low-energy cutoff at energy E$_{0}$, where  
$\xi_{0}$ = E$_{0}$/kT$_{\rm x}$ (eq. [C1] of S02).
We adopt E$_{0}$ = 0.1 keV in this study, as did  S02.
For numerical estimates, the following expression for the primary ionization
rate $\zeta_{\rm x}$ at  distance $R_{\rm X}$ is useful (eq. [3.10] of S02): 

\begin{equation}
\zeta_{\rm x} =  \frac{L_{x}\sigma(kT_{x})}{4 \pi R_{x}^2 kT_{x}}  =  1.13 \times 10^{-8}  \left[{ \frac{L_{x}}{10^{30}~ {\rm erg~ s}^{-1}}} \right] \left[{ \frac{kT_{\rm x}}{{\rm keV}}} \right]^{-(p+1)} \left[{ \frac{R_{\rm x}}{10^{12}~{\rm cm}}} \right]^{-2}~~~{\rm s}^{-1}
\end{equation}
where $\sigma(kT_{x})$ is the photoelectric X-ray absorption cross-section 
per H nucleus evaluated at energy $E = kT_{x}$. 
For a given X-ray photon energy E, the cross-section is approximated by a power-law  

\begin{equation}
\sigma(E) = \sigma(1~  {\rm keV}) \left[{ \frac{E}{1~ {\rm keV}}}  \right]^{-p}~~~{\rm cm^{2}}
\end{equation}
where $\sigma$(1 keV) = 2.27 $\times$ 10$^{-22}$ cm$^{2}$ and p = 2.485
for a solar abundance disk plasma (G97a), as assumed here.
Smaller values of $p$ are required if heavy elements are depleted (G97a).

Following G97a and S02, we position the X-ray source at a 
distance R$_{\rm x}$ = 4R$_{*}$ = 6.4 R$_{\odot}$ = 4.45 $\times$  10$^{11}$  cm
above the disk center. This placement is somewhat arbitrary since existing X-ray
telescopes lack sufficient angular resolution to  pinpoint  the locations of
individual X-ray emitting regions on stars. But, by analogy with the Sun, most
of the emission is thought to originate in coronal loops well above the stellar 
surface. The effects of changes in the location of the X-ray source on 
ionization rates were investigated by IG99 who found that the exact location
has little effect at the low disk column densities of interest here.
We use a  stellar radius R$_{*}$ = 1.6 R$_{\odot}$ appropriate 
to LkCa 15 (Table 1). By comparison,  G97a and S02 considered a generic TTS with 
radius R$_{*}$ = 3 R$_{\odot}$ and  R$_{\rm x}$ $\approx$ 10 - 14 R$_{\odot}$.
The values of $\zeta_{x}$ for the cool (kT$_{x,1}$ = 0.3 keV) and hot
(kT$_{x,2}$ = 5.0  keV) plasma components are given in Table 4 based on the 
X-ray spectral properties of LkCa 15 (model C in Table 3).

The X-ray ionization rate decreases  with radius as $\zeta$ $\propto$ r$^{-2}$ (eq. [1]).
As such, the ionization  rate will be highest in the inner disk close to the
star, all other factors being equal. We  adopt a characteristic radius 
r = 1 AU  $\equiv$ R$_{0}$ in the calculations below, but these results can be readily
scaled to larger radii (Sec. 4.2). To compute the ionization rate at a specific point
(r,z) in the disk, the attenuation  factor $I_{p}(\tau_{x},\xi_{0}$) in eq. (1)
must be calculated. It depends on the X-ray optical depth  given by

\begin{equation}
\tau_{x}(r,z,E) =  \left[{ \frac{r}{R_{\rm x}}}  \right] \sigma({\rm E}) {\rm N}_{\perp,{\rm disk}}(r,z)
\end{equation}
where N$_{\perp,{\rm disk}}$(r,z) is the vertically-integrated column density from infinity  
down to the target height z  above the diskplane (z = 0 at the midplane). 
Specifically,

\begin{equation}
{\rm N}_{\perp,{\rm disk}}(r,z) = \int^{\infty}_{z} {\rm n_{H}(r,\overline{z}) d\overline{z}}~~~{\rm cm^{-2}}
\end{equation}
where n$_{\rm H}$ is the number density of hydrogen nuclei in the disk.
N$_{\perp,{\rm disk}}$ is less than the total column density between the X-ray source and the
target point in the disk according to  the relation
N$_{\perp,{\rm disk}}$(r,z) = (R$_{x}$/r)N$_{\rm H,disk}$(r,z). At r = 1 AU one
obtains R$_{x}$/r $\approx$ 0.03 for the adopted value R$_{x}$ = 
4 R$_{*}$ = 4.5 $\times$  10$^{11}$  cm.

It is obvious from equations (3) and (4) that $\tau_{x}$ for the cool component at a 
given point in the disk will be much larger than the hot component because of
its larger absorption cross-section. In other words, the hard component will
be less-absorbed and penetrate further into the disk. Figure 3 plots $\zeta$ as a 
function of $\tau_{x}$ at r = 1 AU for the cool and hot plasma components.
To generate Figure 3, we have used eqs. (1) and (2) above and have
evaluated  the integral (eq. [C1] of S02)

\begin{equation}
I_{p}(\tau_{x},\xi_{0}) = \int^{\infty}_{\xi_{0}} \xi^{-p}{\rm exp}[-(\xi + \tau_{x} \xi^{-p})]d\xi
\end{equation}
for  $\xi_{0}$ = E$_{0}$/kT$_{x}$
and E$_{0}$ = 0.1 keV using the asymptotic approximation 
$I_{p}(\tau_{x})$ $\sim$ g($\tau_{x}$)$J_{p}(\tau_{x})$
where g($\tau_{x}$) = $\tau_{x}$/($\tau_{0}$ $+$  $\tau_{x}$)
and  $J_{p}(\tau_{x})$ is given by eq. (C3) of S02.
This approximation is accurate for optical depths $\tau_{x}$ $\geq$ $\tau_{0}$
where $\tau_{0}$ = $\xi_{0}^p(p + \xi_{0})p^{-1}$. 
This expression gives $\tau_{0}$  = 6.5 $\times$ 10$^{-2}$
for the kT$_{x}$ = 0.3 keV plasma component ($\xi_{0}$ = 0.33)
and  $\tau_{0}$  = 6.05 $\times$ 10$^{-5}$ for the 
kT$_{x}$ = 5 keV component ($\xi_{0}$ = 0.02). 
The weighting function  g($\tau_{x}$) becomes small for
$\tau_{x}$ $<$  $\tau_{0}$. The effect of the low-energy
cutoff E$_{0}$ is to attenuate the incident X-ray spectrum at 
low energies, as might occur from wind absorption. Raising the
cutoff energy increases  the attenuation. The effect of applying
the cutoff is to flatten the ionization curve at small $\tau_{x}$,
as is shown  for the 0.3 keV plasma component in Figure 3.

To proceed further, we need to evaluate the attenutation
factor $I_{p}(\tau_{x},\xi_{0})$ over a range of X-ray optical 
depths for a specific disk model of LkCa 15. To compute
$\tau_{x}$(r,z,E) we must assume a disk density profile
in order to calculate n$_{\rm H}$(r,z).
We adopt a gas$+$dust disk surface density $\Sigma$(r = 1 AU) $\sim$ 10$^{-3}$ g cm$^{-2}$ 
based on the inner disk model shown in Figure 8 of A11a.
This value, in combination with the 
disk scale height H(r), fixes the mass density $\rho$ at the disk
midplane via the relation
$\rho(r,0)$ = (1/$\sqrt{2 \pi}$)$\Sigma({\rm r})$/${\rm H(r)}$.

The disk mass density scales with radius (r) and height (z)  as
(eq. [5] of IG99)

\begin{equation}
\rho({\rm r,z}) = \rho(r,0) \left[{ \frac{r}{{\rm R_{o}}} } \right]^{q-1.25} {\rm e}^{ {\rm -z^{2}/2H(r)^{2} } }
\end{equation}

In the above, the value of $q$ is determined from the assumed power-law form of
the disk surface density profile $\Sigma$(r) $\propto$ r$^{q}$. Some typical
values used in the literature are $q$ = $-$1.5 (IG99), $q$ = $-$1.0 (A11a),
and $q$ = $-$0.72 (Isella et al. 2012).
To compute the disk scale height H(r = 1 AU) $\equiv$ H$_{0}$ we assume a temperature
T(r = 1 AU) $\approx$ 400 K based on the stellar effective
temperature T$_{*}$ = 4730 K (Table 1) and a radial temperature dependence
T(r) $\propto$ r$^{-0.5}$. Vertical temperature gradients at a given radius
can be ignored (IG99).  The scale height 
H(r) = c$_{s}$/$\sqrt{ {\rm GM_{*}/r^{3}}}$ 
depends on the sound speed
c$_{s}$ = $\sqrt{{\rm (\gamma k T)/(\mu m_{p})}}$, where $\gamma$ is the
adiabatic index, $\mu$ is the mean weight (amu) per particle, and m$_{p}$
is the proton mass. We  have assumed that the hydrogen is mainly molecular at r = 1 AU
($\gamma$ = 1.4,  $\mu$ = 2.3).
If the hydrogen is mainly atomic then the sound speed and scale height 
are a factor of 1.13 larger than for the molecular case. 
The number density of H nuclei follows from the mass density $\rho$ as
n$_{\rm H}$ =  $\rho$/($\mu$m$_{p}$).
Table 4 summarizes the ionization rate at r = 1 AU at the disk midplane
and at one scale height (z = H$_{0}$ = 7.1 $\times$ 10$^{11}$ cm) for 
the cool and hot components.

The  X-ray heating rate per unit volume is proportional to the ionization rate
and is given by (G12)

\begin{equation}
\Gamma_{\rm x} = \zeta {\rm n_{H} Q}
\end{equation}
where $Q$ is the heating rate per ionization. As discussed by G12,
several different processes contribute to the heating rate including
elastic collisions, excitation of rotational and vibrational levels,
H$_{2}$ dissociation, and chemical heating. The heating rate 
depends on the nature of the gas (i.e. atomic versus molecular). 
Assuming that the gas in the LkCa 15 disk at r $\geq$ 1 AU is 
predominantly molecular then the results of G12 for the range of
densities of interest here (n$_{\rm H}$ $\sim$ 10$^{5}$ -
10$^{8}$ cm$^{-3}$; Table 4)  give $Q \approx$ 15 - 18 eV.
Representative heating rates $\Gamma_{\rm x}$ are given in Table 4
based on an assumed value $Q$ = 17 eV.

\subsection{Scaling Relations}

The above results can be extrapolated to larger radii using scaling
relations: T(r) $\propto$ r$^{-0.5}$, H(r) $\propto$ r$^{+1.25}$
and $\Sigma$(r) $\propto$ r$^{q}$. The value of  $q$ is in principle  constrained 
by observations but its value in the inner disk region of LkCa 15 is not well-determined
because of the inability of current (sub)millimeter  telescopes to resolve the inner
disk close to the star.  The improved resolution now becoming  available with ALMA
offers the possibility of placing  tighter observational constraints on conditions 
in the inner disk.  The mass density
scales as $\rho$(r) $\propto$ r$^{q-1.25}$ as does n$_{\rm H}$(r). 
As a representative case we take $q$ = $-$1.0, as for the 
inner disk model of A11a. Table 4 gives the ionization and heating rates
in the disk midplane at the protoplanet distance r = 15 AU for
the case $q$ = $-$1.0. Note that  the vertical column density scales as
N$_{\perp,{\rm disk}}$(r) $\propto$ r$^{q}$, so the value $q$ = $-$1.0 results in
a fortuitous cancellation that makes  $\tau_{x}$(r)  independent 
of $r$ (eq. (3). The heating rate per unit volume 
$\Gamma_{\rm x}$ = $\zeta$n$_{\rm H}$$Q$ is easily scaled, 
being linear in all three parameters.

\subsection{X-rays in the Low-Density Transition Disk}

A few comments on the specific LKCa 15 case  considered above 
are useful to compare with the analyses of generic TTS by
G97a, S02, and IG99. These previous studies
adopted a disk model based on properties of the minimum mass solar 
nebula with a surface density $\Sigma$(r = 1 AU) $\sim$ 10$^{3}$ g cm$^{-2}$.
In sharp contrast, the inner region of the disk surrounding LkCa 15 has a
surface density $\Sigma$(r = 1 AU) $~\sim$ 10$^{-3}$
g cm$^{-2}$, about 6 orders of magnitude lower than that of the minimum
mass solar nebula (Fig. 8 of A11a). Consequently, the 
number density n$_{\rm H}$ at r = 1 AU is much less, being of order
n$_{\rm H}$ $\sim$  10$^{8}$ cm$^{-3}$ near the midplane (Table 4).

The low-density environment of the inner disk of LkCa 15 has two important 
consequences. First, the X-ray optical depth at a given radius and  energy  
is much less than would be the case in the minimum mass 
solar nebula, allowing X-rays to penetrate deeper into the 
inner regions of the LkCa 15  disk. As Table 4 and Figure 3 show, the hard X-ray
component has $\tau_{x}$ $\approx$ 0.02 at the midplane (r = 1 AU) so most of the
incident flux of E = 5 keV photons at r = 1 AU penetrates all the way to the midplane.
At such low optical depths, the hard photons can even pass through the entire
disk without being absorbed.
By contrast, the soft component has  $\tau_{x}$ $\approx$ 20 at the midplane
so  only a small fraction of the incident E = 0.3 keV photons reaches the midplane.
However,  $\tau_{x}$ = 1 for the cool 0.3 keV  component occurs at
z = 1.95 H$_{0}$  (r = 1 AU) and even at z = H$_{0}$ the 
ionization rates of the cool and hot components are nearly the same
(Fig. 3 and Table 4). Thus, in the higher disk layers the effects
of the cool component are not negligible.
Second, the heating rate (eq. 8) is lower in the low-density inner
disk of LkCa 15 than in denser disk environments 
because  there are fewer atoms to absorb  the X-ray photons. Since
n$_{\rm H}$ $\propto$ r$^{-2.25}$ (assuming $q$ = $-$1.0), 
the particle density falls off rapidly and the X-ray heating rate is
quite low at  the distance of the protoplanet r $\approx$ 15 AU (Table 4).

\section{Summary}

The {\em Chandra} X-ray data  analyzed here reveal that
LkCa 15 is a more luminous X-ray source than previously 
assumed on the basis of the {\em ROSAT} All-Sky Survey non-detection 
and  is capable of significantly  
influencing physical conditions in the inner low-density region of 
the transition disk. The X-ray emission is described by
a  cool component at temperature
kT$_{1}$ $\approx$ 0.3 keV and a hotter  component
at kT$_{2}$ $\approx$ 5 keV. Although the harder 5 keV component
is able to penetrate further into the disk, the effects of
the cool component must also be taken into account at higher
disk layers z $>$ H$_{0}$. The X-ray properties determined here 
provide valuable input data for more sophisticated models of
the LkCa 15 disk based on radiative transfer calculations
(e.g. IG99; Nomura et al. 2007). Such detailed models would need
to consider  effects such as X-ray scattering, possible
abundance  variations within the disk (G97a), UV radiation, 
mechanical and viscous heating, and cosmic rays. Previous
studies (IG99; Dolginov \& Stepinski 1994) suggest that cosmic ray 
ionization will be  less important than X-ray ionization at the  
relatively low midplane vertical column densities  N$_{\perp \rm ,disk}$ $\sim$ 10$^{20}$ cm$^{-2}$
estimated for  the LkCa 15 transition disk.  In addition, the short {\em Chandra}
observation discussed here needs to be followed up  
by further time monitoring to determine if LkCa 15 undergoes
large X-ray flares. If so, the X-ray luminosity and temperature
of the hot component could significantly exceed the values
adopted here for brief periods of time ($\sim$hours to days), 
leading to punctuated increases in  ionization and heating rates.

\acknowledgments

This work was supported by NASA GSFC award NNG05GE69G.
The Chandra X-ray Observatory Center (CXC) is operated by the 
Smithsonian Astrophysical Observatory (SAO) for, and on behalf of, 
the National Aeronautics Space Administration under contract NAS8-03060.

%
%


\clearpage

\begin{deluxetable}{ccccccccccc}
\tabletypesize{\scriptsize}
\tablewidth{0pt}
\tablecaption{Properties of LkCa 15}
\tablehead{
           \colhead{Type}               &
           \colhead{Age}                &
           \colhead{M$_{*}$}            &
           \colhead{R$_{*}$}            &
           \colhead{T$_{eff}$}          &
           \colhead{L$_{*}$}            &
           \colhead{M$_{disk}$}         &
           \colhead{$\dot{\rm M}_{acc}$} &
           \colhead{$i_{disk}$}         &
           \colhead{A$_{V}$}            &
           \colhead{d}                \\
           \colhead{}                   &
           \colhead{(My)}                   &
           \colhead{(M$_{\odot}$)}                   &
           \colhead{(R$_{\odot}$)}                   &
           \colhead{(K)}                   &
           \colhead{(L$_{\odot}$)}                   &
           \colhead{(M$_{\odot}$)}                    &
           \colhead{(M$_{\odot}$/yr)}                  &
           \colhead{(deg.)}                   &
           \colhead{(mag)}                   &
           \colhead{(pc)}                   
}
\startdata
 K5   & 2 [1 - 4]      &  1.0           &  1.6          & 4730       & 0.74 - 1.2    & 0.055          & 2 $\times$ 10$^{-9}$&  51        & 1.3 - 1.7    & 140    \\
\enddata
\tablecomments{Data are from Kenyon \& Hartmann  1995; Simon et al. 2000;
           Pi\'{e}tu et al. 2007; Espaillat et al. 2010;
           Andrews et al. 2011a,b;
           Kraus \& Hillenbrand 2009; Isella et al. 2012; Kraus \& Ireland 2012.}
\end{deluxetable}

\clearpage

\begin{deluxetable}{lllccccl}
\tabletypesize{\scriptsize}
\tablewidth{0pt} 
\tablecaption{ X-ray Properties of LkCa 15 (Chandra ACIS-S)}
\tablehead{
	 \colhead{Name}	&
           \colhead{R.A.}               &
           \colhead{decl.}              &
           \colhead{Net Counts}         &
           \colhead{E$_{50}$}           &
           \colhead{P$_{const}$}             &
           \colhead{log L$_{x}$}             &
           \colhead{Identification(offset)}      \\    
           \colhead{}	&
           \colhead{(J2000)}                 &
           \colhead{(J2000)} &
           \colhead{(cts)}                                          &               
           \colhead{(keV)}                                          &                  
           \colhead{}                                          & 
           \colhead{(ergs s$^{-1}$)}                                          &
           \colhead{(arcsec)} 
                                  }
\startdata
LkCa 15 	& 04 39 17.793 & $+$22 21 03.28 & 590 $\pm$ 24 & 1.01 & 0.91 & 30.4 & GSC J043917.787$+$222103.26 (0.08) \\
\enddata
\tablecomments{
The nominal pointing position for
the observation was (J2000.0) RA = 04$^h$ 39$^m$ 18.52$^s$,
decl. = $+$22$^{\circ}$ 20$'$ 49$''$.1, which lies
17$''$.5 SE of  LkCa 15.
X-ray data are from CCD7 (ACIS chip S3) using events in the 0.2 - 8 keV range inside a 
circular source extraction region of radius 1$''$.5. 
Tabulated quantities are: J2000.0 X-ray position (R.A., decl.), total source counts accumulated
in a 9813 s exposure,  median photon  energy (E$_{50}$),
probability of constant count-rate determined by the Gregory-Loredo 
algorithm (P$_{const}$); unabsorbed X-ray luminosity (0.3 - 10 keV; see also Table 3),
and {\em HST} GSC v2.3.2  counterpart identification.
The offset (in parenthesis) is given in arc seconds between the X-ray and GSC counterpart position.}

\end{deluxetable}

\clearpage

\begin{deluxetable}{llll}
\tabletypesize{\scriptsize}
\tablewidth{0pc}
\tablecaption{{\em Chandra} Spectral Fits for LkCa 15
   \label{tbl-1}}
\tablehead{
\colhead{Parameter}      &
\colhead{ }              &
\colhead{  }
}
\startdata
Model\tablenotemark{a}                  &           A           &    B                        & C               \nl
Emission                                & Thermal (2T)          &  Thermal (2T)               & Thermal (2T)    \nl
Abundances                              & solar\tablenotemark{b}&  non-solar\tablenotemark{c} & non-solar\tablenotemark{d}    \nl
N$_{\rm H}$ (10$^{22}$ cm$^{-2}$)       & 0.77 [0.63 - 0.88]    & 0.76 [0.57 - 0.87]          & 0.37 [0.24 - 0.51] \nl
kT$_{1}$ (keV)                          & 0.18 [0.15 - 0.24]    & 0.18 [0.15 - 0.24]          & 0.30 [0.25 - 0.37] \nl
kT$_{2}$ (keV)                          & 4.86 [3.16 - 10.5]    & 4.25 [2.65 - 9.00]          & 5.11 [3.00 - 13.6] \nl
norm$_{1}$ (10$^{-2}$)\tablenotemark{b} & 0.67 [0.17 - 2.04]   & 1.28 [0.26 - 6.80]           & 0.08 [0.04 - 0.19] \nl
norm$_{2}$  (10$^{-4}$)\tablenotemark{b}& 1.55 [1.24 - 1.86]    & 1.82 [1.09 - 2.56]          & 1.58  [1.23 - 2.04] \nl
Z (Z$_{\odot}$)                         & \{1.0\}                 & 0.46 [0.13 - ...]         & ...                 \nl
$\chi^2$/dof                            & 44.2/42               & 43.4/41                     & 41.9/42             \nl
$\chi^2_{red}$                          & 1.05                  & 1.06                        & 1.00                \nl
F$_{\rm X}$ (10$^{-12}$ ergs cm$^{-2}$ s$^{-1}$)   & 0.34 (11.6)& 0.32 (10.9)                 & 0.32 (1.07)         \nl
F$_{\rm X,1}$ (10$^{-12}$ ergs cm$^{-2}$ s$^{-1}$) & 0.15 (11.3)& 0.15 (10.6)                 & 0.18 (0.25)         \nl
F$_{\rm X,2}$ (10$^{-12}$ ergs cm$^{-2}$ s$^{-1}$) & 0.19 (0.30)& 0.17 (0.30)                 & 0.14 (0.81)         \nl
log L$_{\rm X}$ (ergs s$^{-1}$)                    & 31.43      & 31.41                       & 30.40               \nl
log L$_{\rm X,1}$ (ergs s$^{-1}$)                  & 31.42      & 31.39                       & 29.77               \nl
log L$_{\rm X,2}$ (ergs s$^{-1}$)                  & 29.85      & 29.85                       & 30.28               \nl
log [L$_{\rm X}$/L$_{bol}$]                        & $-$2.15    & $-$2.18                     & $-$3.18             \nl
\enddata
\tablecomments{
Based on  XSPEC (vers. 12.7.1) fits of the background-subtracted ACIS-S spectrum binned
to a minimum of 10 counts per bin using 9,813 s of  exposure time. The fits for models
A and B were obtained using an absorbed  $apec$ optically plasma model and model C
used a similar absorbed  $vapec$ model that allows the  abundance of each element
to be specified. The tabulated parameters
are absorption column density (N$_{\rm H}$), plasma energy (kT),
and XSPEC component normalization (norm).
Solar abundances are referenced to  Anders \& Grevesse (1989).
Square brackets enclose 90\% confidence intervals and an ellipsis means that
the algorithm used to compute confidence intervals did not converge.
Quantities enclosed in curly braces were held fixed during fitting.
The total X-ray flux (F$_{\rm X}$) and fluxes associated with each model component
(F$_{\rm X,i}$)  are the absorbed values in the 0.3 - 10 keV range, followed in
parentheses by  unabsorbed values.
The total X-ray luminosity L$_{\rm X}$  and luminosities of each component
L$_{\rm X,i}$ are unabsorbed values in the 0.3 - 10 keV range and  assume a
distance of 140 pc. A value L$_{bol}$ = 1.0 L$_{\odot}$ is adopted
based on an average of values given in the literature.}
\tablenotetext{a}{Models A, B, and C are of form:~N$_{\rm H}$$\cdot$(kT$_{1}$ $+$ kT$_{2}$)} \\
\tablenotetext{b}{For thermal $apec$ models, the norm is related to the volume emission measure
                  (EM = n$_{e}^{2}$V)  by
                  EM = 4$\pi$10$^{14}$d$_{cm}^2$$\times$norm, where d$_{cm}$ is the stellar
                  distance in cm. At d = 140 pc, this becomes
                  EM = 2.34$\times$10$^{56}$ $\times$ norm (cm$^{-3}$). }
\tablenotetext{c}{The global metallicity Z was allowed to vary. Z = 1.0 corresponds to solar abundances.}
\tablenotetext{d}{Abundances were held fixed at typical values for TTS in Taurus (G\"{u}del et al. 2007; Scelsi et al. 2007). 
                  These are (relative to solar): H = 1.0, He = 1.0, C = 0.45, N = 0.79, O = 0.43,
                   Ne = 0.83, Mg = 0.26, Al = 0.50, Si = 0.31, S = 0.42, Ar = 0.55, Ca = 0.195, 
                   Fe = 0.195, Ne = 0.195. }
\end{deluxetable}

\clearpage

\begin{deluxetable}{cccclllc}
\tabletypesize{\footnotesize}
\tablewidth{0pt}
\tablecaption{ X-ray Ionization and Heating Rates of LkCa 15}
\tablehead{
           \colhead{r}               &
           \colhead{z/H(r)}              &
           \colhead{n$_{\rm H}$}         &
           \colhead{E}           &
           \colhead{$\tau_{x}$}             &
           \colhead{$\zeta_{x}$}             &
           \colhead{$\zeta$}             &
           \colhead{$\Gamma_{x}$}      \\
           \colhead{(AU)}   &
           \colhead{}                 &
           \colhead{(cm$^{-3}$)} &
           \colhead{(keV)}                                          &
           \colhead{}                                          &
           \colhead{(s$^{-1}$)}                             &
           \colhead{(s$^{-1}$)}                             &
           \colhead{(ergs s$^{-1}$ cm$^{-3}$)}
}
\startdata
  1 & 0 & 1.46e8 & 0.3 & 19.8    & 2.23e-06  & 2.87e-11   & 1.14e-13  \\
  1 & 0 & 1.46e8 & 5.0 & 0.018   & 3.97e-10  & 1.49e-10   & 5.92e-13  \\
  1 & 1 & 0.88e8 & 0.3 & 6.3     & 2.23e-06  & 1.76e-10   & 4.21e-13  \\
  1 & 1 & 0.88e8 & 5.0 & 0.006   & 3.97e-10  & 3.98e-10   & 9.53e-13  \\
 15 & 0 & 3.30e5 & 0.3 & 19.8    & 2.23e-06  & 1.28e-13   & 1.15e-18   \\
 15 & 0 & 3.30e5 & 5.0 & 0.018   & 3.97e-10  & 6.63e-13   & 5.95e-18   \\
\enddata
\tablecomments{
Based on inner disk model discussed in text with surface density
profile $\Sigma$(r) $\propto$ r$^{-1}$ and 
$\Sigma$(r = 1 AU) = 10$^{-3}$ g cm$^{-2}$. 
Disk scale heights are H(r = 1 AU)  = 7.1 $\times$ 10$^{11}$ cm  
and H(r = 15 AU)  = 2.1 $\times$ 10$^{13}$ cm assuming molecular
hydrogen gas and temperatures T(r = 1 AU) $\approx$ 400 K
and T(r = 15 AU) $\approx$ 100 K.
Solar abundances are assumed in the disk. X-ray luminosities
and temperatures used to compute the primary ionization rate $\zeta_{x}$
are given in Table 3 (model C). The X-ray source is assumed to be located
at a height of 4 R$_{*}$ = 6.4 R$_{\odot}$ above the disk center.
The heating rate $\Gamma_{\rm x}$ = $\zeta$n$_{\rm H}$$Q$ assumes
Q = 17 eV but the value of $Q$ is dependent on gas properties (G12).
}

\end{deluxetable}
\clearpage


\begin{figure}
\figurenum{1}
\epsscale{1.0}
\includegraphics*[width=8.0cm,angle=-90]{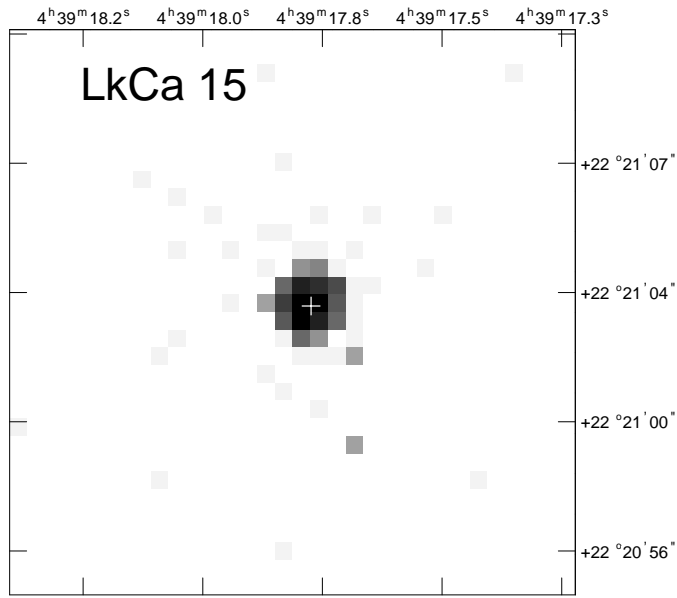}

\caption{Broad-band (0.2 - 8 keV)  ACIS-S image of LkCa 15. 
The $+$ sign marks the {\em HST} GSC v2.3.2  position of LkCa 15.
Pixel size = 0$''$492; log intensity scale.
}
\end{figure}

\clearpage

\begin{figure}
\figurenum{2}
\epsscale{1.0}
\includegraphics*[width=8.0cm,angle=-90]{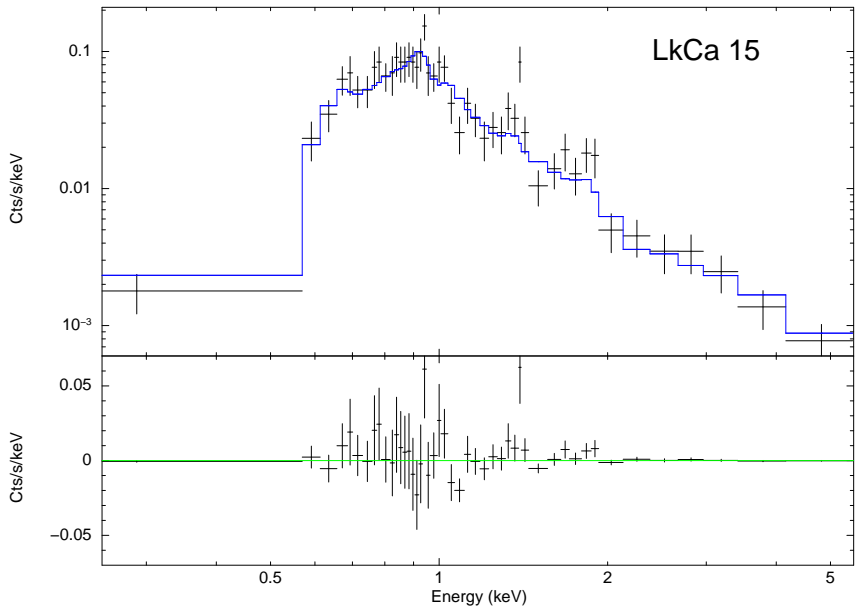} 
\caption{
Chandra ACIS-S spectrum of LkCa 15 binned to a minimum of 10 counts per bin.
Solid line in upper panel shows the best-fit 2T $vapec$ model with non-solar
abundances (model C in Table 3). Bottom panel shows fit residuals.
}
\end{figure}

\clearpage

\begin{figure}
\figurenum{3}
\epsscale{1.0}
\includegraphics*[width=10.5cm,angle=-90]{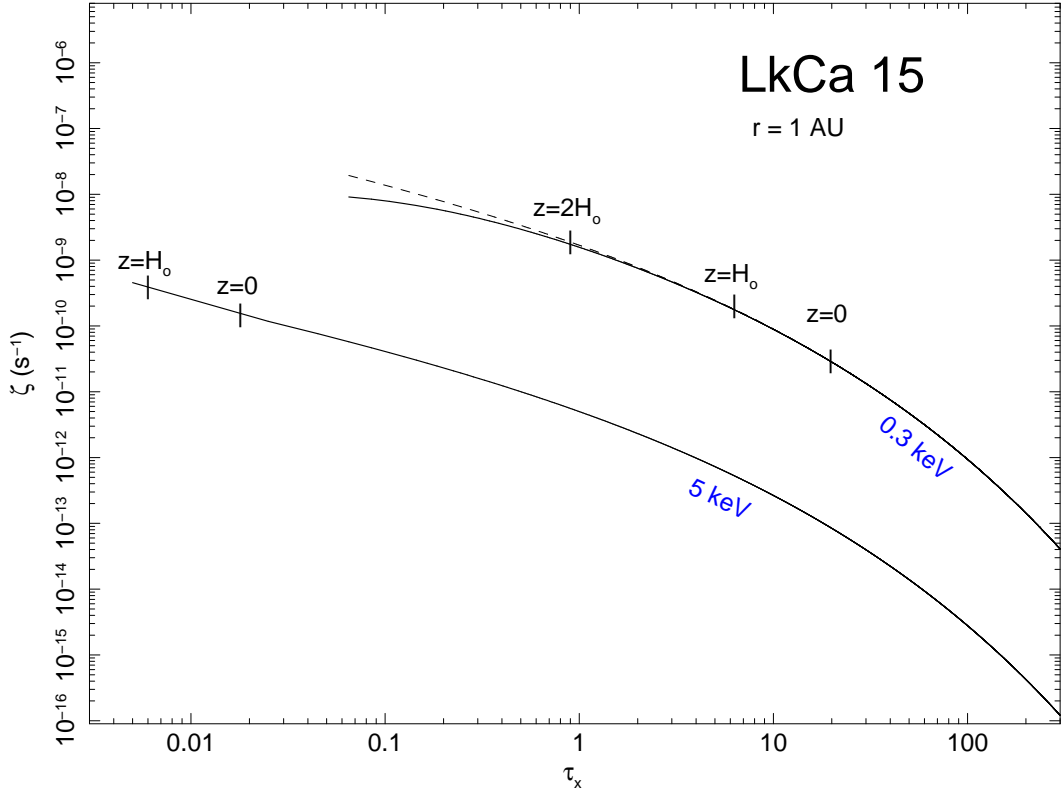}
\caption{Ionization rate for the cool and hot plasma components
as a function of X-ray optical depth for LkCa 15 at a radial 
distance r = 1 AU. The curves are based on the disk model discussed in the 
text (Sec. 4.1) and values of kT and L$_{x}$ determined from
the X-ray spectral fit in model C of Table 3. The solid lines assume
a low-energy cutoff E$_{0}$ = 0.1 keV and the dashed line shows a less 
stringent cutoff E$_{0}$  = 0.01 keV for the cool component. The curve
for 0.3 keV is truncated at $\tau_{x}$ = 0.065 since the expression
used to compute the attenuation is not valid at smaller values of $\tau_{x}$.
The short vertical lines mark the optical depth at the disk midplane
(z = 0) and at multiples of one  scale height (H$_{0}$) for 
each component (Table 4). At r = 1 AU, the vertically-integrated column
density at the midplane is N$_{\perp,{\rm disk}}$(r = 1 AU, z=0) =
1.3 $\times$ 10$^{20}$ cm$^{-2}$.
}

\end{figure}

\clearpage

\end{document}